\documentclass[twocolumn,pre,a4paper,superscriptaddress,showpacs,showkey,nofootinbib]{revtex4}
\usepackage{amsfonts,amsmath,amssymb,graphicx,epsfig}

\usepackage{color}

\newcommand{\rcorrec}[1]{\textcolor{blue}{#1}}

\newcommand{\previous}[1]{}
\newcommand{\comment}[1]{}

\def\chemin{./}

\newcommand{\pic}[3]{\includegraphics[clip=true,width=#2 \linewidth,angle=#3]{\chemin #1}}

\newcommand{\figcap}[2]{\caption{\label{#1}\em #2}}

\newcommand{\bmini}[1]{\begin{minipage}{#1 \linewidth}}
\newcommand{\emini}{\end{minipage}}

\begin{document}
\draft
\title{Local waiting time fluctuations along a randomly pinned crack front}
\author{Knut J\o rgen M\aa l\o y}
\author{St\'ephane Santucci}
\affiliation{Fysisk Institutt, Universitetet i Oslo,
       P. O. Boks 1048 Blindern, N-0316 Oslo 3, Norway}
\author{Jean Schmittbuhl}
\author{Renaud  Toussaint}
\affiliation{Institut de Physique du Globe de Strasbourg, UMR 7516,\\ 5 rue Ren\'e Descartes, F-67084 Strasbourg Cedex, France.}

\begin{abstract}
The propagation of an interfacial crack along a heterogeneous weak
plane of a transparent Plexiglas block is followed  using a
high resolution fast camera. We show that the fracture front dynamics
is governed by local and irregular avalanches with very large size and
velocity fluctuations. We characterize the intermittent dynamics  observed, i.e.
the local pinnings and depinnings of the crack front
which trigger a rich burst activity, by measuring the local waiting
time fluctuations along the crack front during its propagation.  The
local front line velocity distribution deduced from the waiting time
analysis exhibits a power law behavior, 
$P(v) \propto v^{-\eta}$ with $\eta = 2.55 \pm 0.15$, for velocities $v$ larger than the average
front speed $\langle v \rangle$. The burst size distribution is also a power law,
 $P(S)\propto S^{-\gamma}$ with $\gamma=1.7 \pm 0.1$.
Above a characteristic length scale of disorder $L_d \sim 15
\mu m$, the avalanche clusters become anisotropic, and the
scaling  of the anisotropy ratio
provides an estimate of the roughness exponent of
the crack front line, $H=0.66$, in close agreement with previous
independent estimates.
\end{abstract}
 \maketitle


The physics community has recently paid a lot of attention to the
study of damaging processes
\cite{celarie03,hansen03,afek05}. This interest is motivated
not only by the practical benefits to many engineering domains, but
also from a more fundamental point of view, by the diverse challenging
questions brought forward, in particular in statistical physics
\cite{herrmann90}.  The role of heterogeneities during crack
propagation is of central importance since they induce local pinnings
of the crack front and subsequently trigger a very complex history of
the fracture in the material. One of the consequences of this
phenomenology is the roughness of fracture surfaces left by the
crack.  
Indeed, cracks in heterogeneous media
exhibit a self-affine morphology, with long
range correlations. The associated roughness exponent was found to be very
robust for different materials, over a broad range of length scales
\cite{mandelbrot84,brown85,bouchaud90,maloy92,schmittbuhl93,schmittbuhl95,bouchaud97}, 
and was further conjectured to be universal \cite{bouchaud90,maloy92}.
A recent work
\cite{hansen03,schmittbuhl03} suggests that the origin of these
self-affine long range correlations comes from the elastic
interactions within the damage zone and proposes a link between the
roughness exponent and the critical exponent $\nu$ for the correlation
length of the damage clusters.  More generally, front
propagation in random media has become a challenging problem related to 
the dynamics of interfaces in many different physical
systems theoretically connected, such as crack fronts \cite{bouchaud97}, magnetic domain walls
\cite{Lemerle98}, or wetting contact lines
\cite{Prevost02,Moulinet02,Moulinet04}, where elasticity and disorder
compete to shape the interface.

In order to shed some light on the interactions between the crack
front and material heterogeneities, a simplification to a two
dimensional configuration -an interfacial crack- has been proposed
both experimentally \cite{schmittbuhl97,delaplace99} and theoretically
\cite{schmittbuhl95b,schmittbuhl03}. 
The interfacial configuration provides a higher resolution since
all locations of the crack front belong to the same plane.
Moreover, using a
transparent material and a high resolution fast camera, the detailed
complex crack dynamics can be captured, following the crack front
with a high precision both in time and space \cite{maloy01}.
So far experiments have been focused on  
the fracture front line morphology leading to the estimated roughness
exponent $\zeta=0.55 \pm 0.03$ \cite{schmittbuhl97}, followed up by a
longer study showing $\zeta=0.63 \pm 0.03$ \cite{delaplace99}.  First
attempts have been recently performed to \previous{analyse} analyze the interfacial crack
front dynamics \cite{maloy01,santucci05}. These studies have shown
that the fracture front propagation is intermittent and can be described in 
terms of a Family-Vicsek scaling \cite {family85} with a 
roughness $\zeta=0.6$ and a dynamic exponent $\kappa=1.2 \pm 0.2$.

In this Letter, we study a system first 
studied experimentally by  Schmittbuhl and M\aa l\o y \cite{schmittbuhl97, maloy01}.
Whereas previous studies focused on the morphology of the interfacial crack \cite{schmittbuhl97}, we
focus on the local crack dynamics, and on the distribution in both time and space of the waiting 
time during pinning events. To address this problem, we introduce a new analysis procedure in 
order to study the local waiting time fluctations. The improved experimental techniques and 
resolution allow to show that the dynamics of the fracture front is driven by local 
irregular avalanches with very large size and velocity fluctuations, and anisotropic 
shapes whose scaling is directly linked to the self-affine scaling of the crack 
front itself. This new set of experiments also confirms earlier results on such 
systems \cite{schmittbuhl97,maloy01}.

We describe here experiments where two Plexiglas plates are annealed
together to create a single block with a weak interface
\cite{schmittbuhl97}.  The plates are of dimensions: $32cm\times 14cm
\times 1cm$ and $34cm\times 12cm \times 0.4cm$, and annealed together
at $205^{\circ}$C under several bars of normal pressure.  Before
annealing, both plates are sand-blasted on one side with $50 \mu m$
steel particles or $100 \mu m$ glass beads.  Sand-blasting introduces
a random topography which induces local toughness fluctuations during
the annealing procedure. We have measured the profile of a
sand-blasted Plexiglas surface, using a white light interferometry
technique (performed at SINTEF Oslo laboratory) and found that the
local irregularities have a characteristic size about $ 15 \mu
m$ \cite{santucci05b}. While the upper Plexiglas plate is
clamped to a stiff aluminium frame, a press applies a normal
displacement to the lower one (1cm thick) at a constant low speed which results in a stable
crack propagation in mode I \cite{schmittbuhl97}. The fracture front is observed with a high
resolution fast camera mounted on a microscope.  Two different cameras
have been used, a Kodak Motion Korder Analyzer CCD camera
which records up to $500$ frames per second (fps) with a 512x240 pixel
resolution, and lately a much more powerful one, a Photron Ultima CMOS
camera. Using this camera at a spatial resolution of $1024 \times 512$
pixels, and an acquisition rate of $1000$ fps we can follow the stable
crack front during more than $12s$ (recording up to $ 12288$
images). Different experiments have been performed varying
the acquisition hardware, the microscope magnification
corresponding to a pixel size  between $1.7 $ to $ 10\mu m$, and the
average front line speed ranging from $0.35 $ to $40 \mu m/s$.  It is
important to note that in all cases, the pixel size is smaller than
the size of the local irregularities of about $20 \mu m$ due to the
sand-blasting process.
\begin{figure}
\pic{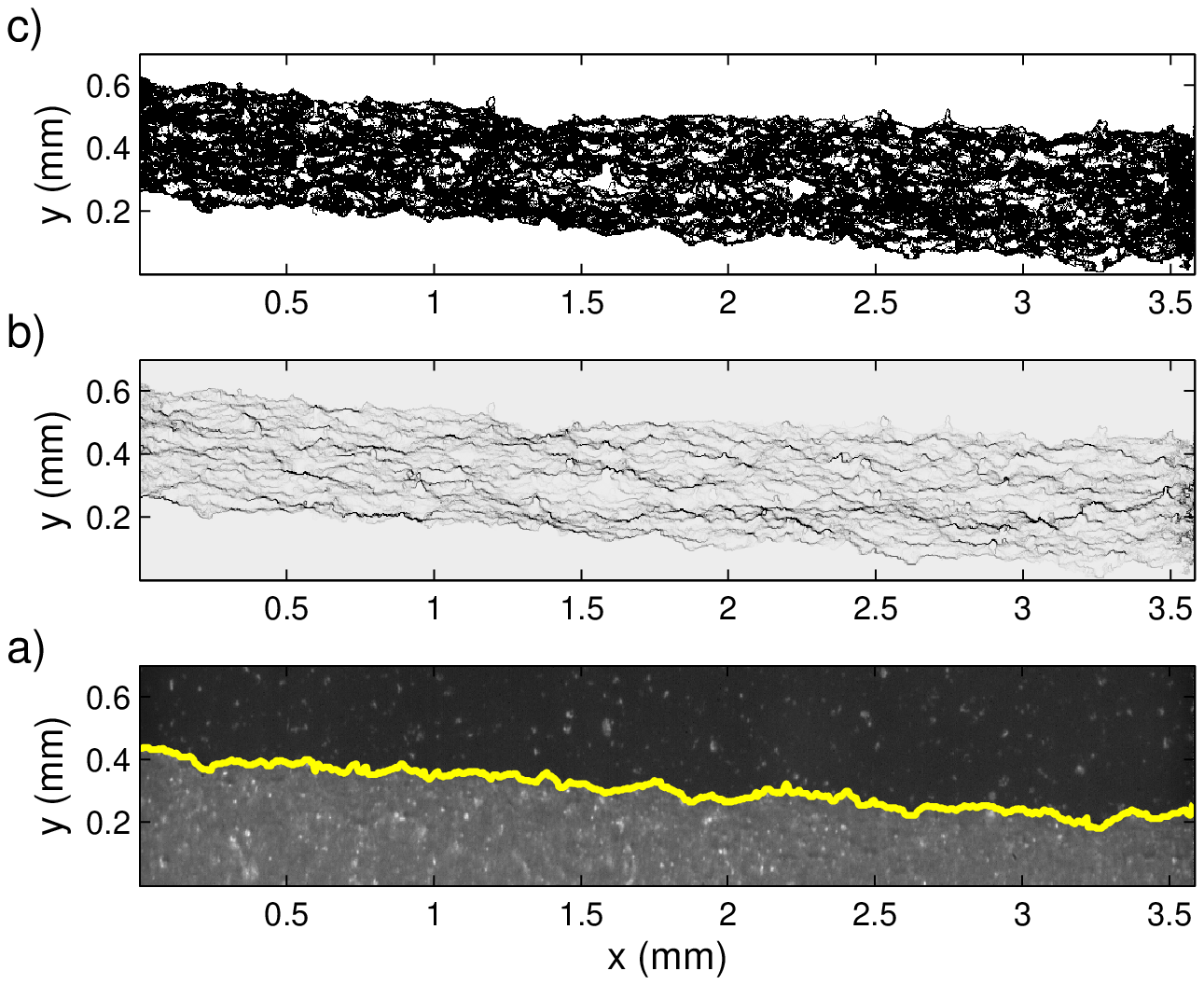}{1.0}{0}\figcap{Fig1}{ $a)$ Typical example of a
picture recorded by the high speed camera (Photron Ultima) during an
experiment with an average crack front speed $\langle v \rangle =28.1
\mu m.s^{-1}$, and a pixel size $a=3.5\mu m$. The solid line
represents the interface separating the uncracked (in black) and
cracked parts extracted after image analysis.  b) Gray scale map of
the waiting time matrix deduced from $10000$ front positions recorded
at a rate of $1000$ fps. The darker parts, the longer waiting times.
c) Spatial distribution of clusters (in white) corresponding to
velocities ten times larger than the average crack front
speed.}\label{Fig1}
\end{figure}

In order to  analyze the local waiting time fluctuations and 
the burst dynamics, we propose the following procedure: the fracture front
lines extracted from image analysis of the digital pictures (see
Fig.~\ref{Fig1}a) are added to obtain a waiting time matrix ${\bf
W(x,y)}$.
 This matrix has the dimension of the original image and an
initial value equal to zero. We add the value 1 to the matrix element
$w$ corresponding to each pixel of the detected front line
position $(x,y)$. This procedure is performed for all frames of a given
experiment in order to obtain the final waiting time matrix ${\bf
W(x,y)}$.  A gray scale map of this matrix is shown in Fig.~\ref{Fig1}b.
The spatially random toughness along the weak
interface generates a rough crack line in pinning the crack front
(Fig.~\ref{Fig1}a), and triggers a rich burst activity on a wide range
of length scales. The numerous and various regions of gray levels
suggest this intermittent dynamics (Fig.~\ref{Fig1}b).  It is
important to mention that the image recording is so fast that there
are basically no holes in the waiting time matrix ${\bf
W(x,y)}$, {\it i.e.} no regions of zero values (apart from below
the first front, above the last one, and a few artifacts due to
impurities in the sample).  Then, we can deduce from ${\bf W(x,y)}$, a
matrix ${\bf V(x,y)}$ of the local normal speed of the interface at
the time when the front went through a particular position, by
computing the inverse value of the corresponding matrix element $w$ of
${\bf W(x,y)}$ multiplied by the ratio of the pixel size $a$
and the typical time between two images $\delta t$.
Therefore, we can associate to each pixel corresponding to the crack
line in each image, a local front velocity $v =
\frac{1}{w}\frac{a}{\delta t}$.  Finally, we can obtain the
probability distribution functions of the local waiting time $w$ and
the local front velocity $v$, by estimating the occurrence number of
each measured waiting time or velocity on all the pixels in all the
fracture front line images.
\begin{figure}
 \pic{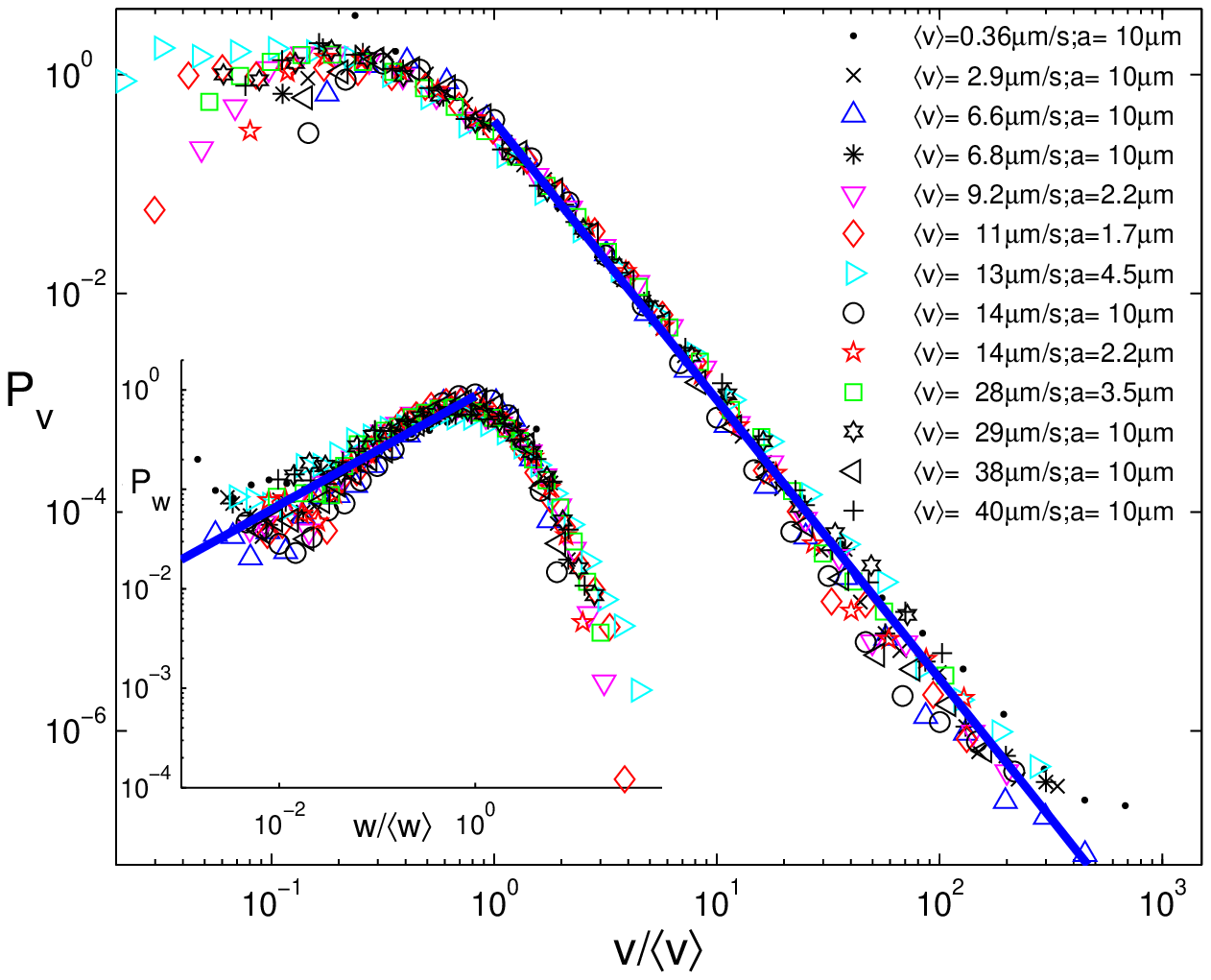}{1.0}{0} \figcap{Fig2}{The velocity distribution
 $P(v/\langle v \rangle)$ as a function of the scaled velocity
 $v/\langle v \rangle$ for different experimental conditions 
 (various average crack front speeds $\langle v \rangle$ and pixel sizes $a$). 
 A fit (solid line) to all data for $v> \langle v \rangle$ has a slope
 $-2.55$. Inset shows the corresponding waiting time distribution 
 $P(w/\langle w \rangle)$ as function of the scaled waiting time $w/\langle w \rangle$.  The solid line represents a 
 fit to all the data for $ w < \langle w \rangle$ with 
 a slope $0.55$.}\label{Fig2}
\end{figure}
The velocity distribution $P(v/\langle v \rangle)$ is shown in Fig.~\ref{Fig2}
in a log-log scale. 
A data collapse is obtained for all different experimental conditions
by scaling the local velocity $v$ with the average crack front speed
$\langle v \rangle $ which varies from one experiment to
another. A clear power law behavior of the velocity distribution
$P(v/\langle v \rangle) \sim (v/\langle v \rangle)^{-\eta}$ is
observed for velocities larger than $\langle v \rangle$ with a
crossover to a slowly increasing function for velocities smaller than
$\langle v \rangle$.  A linear fit to the experimental data for
$v/\langle v \rangle > 1$ gives a slope $-\eta=-2.55 \pm 0.15$.  The
inserted figure shows a double logarithmic plot of the corresponding
waiting time distribution $P(w/\langle w \rangle)$, where $w$ is the
waiting time, and $\langle w \rangle$ the average waiting time for
each experiment.  A linear fit to the experimental data for $w<
\langle w \rangle$ has a slope $0.55 \pm 0.15$ consistent with the
exponent $\eta-2$ deduced from the velocity distribution. The power
law distribution of the local velocities confirms once again previous
observations, revealing a non trivial underlying dynamics as observed
on a fast video recording.
It is important to note that even though the first moment of the
velocity fluctuations $ \langle v \rangle $ exists, the second and
higher moments are ill defined and dominated by the largest velocity
fluctuations. 
In an earlier work, the velocity distribution was investigated with a different method, 
based on the distance between subsequent fronts at a given time interval \cite{maloy01}. 
However, such method proved out to produce results depending on the time interval chosen. Indeed, a
short time between the front only gave contribution from the high
velocity part of the distribution while a long time between the fronts
gave a peak around the average velocity only.  By using the
concept of  waiting time introduced in the present Letter, we are able to
measure both high and low velocities.  In the present case there exists a well 
defined length scale $a$ at
which the velocity can be measured.
Using  different magnification of the microscope, we have checked the
robustness of our procedure and shown the reproducibility of our
results for different pixel sizes.

In order to analyze the local burst activity, we consider a thresholded matrix
generated from the velocity matrix ${\bf V(x,y)}$, by setting the matrix 
elements $v$ equal to one for $v>C\cdot \langle v \rangle $ and zero elsewhere,
where $C$ is a constant of the order of a few unities. 
\begin{figure}[h!]
 \pic{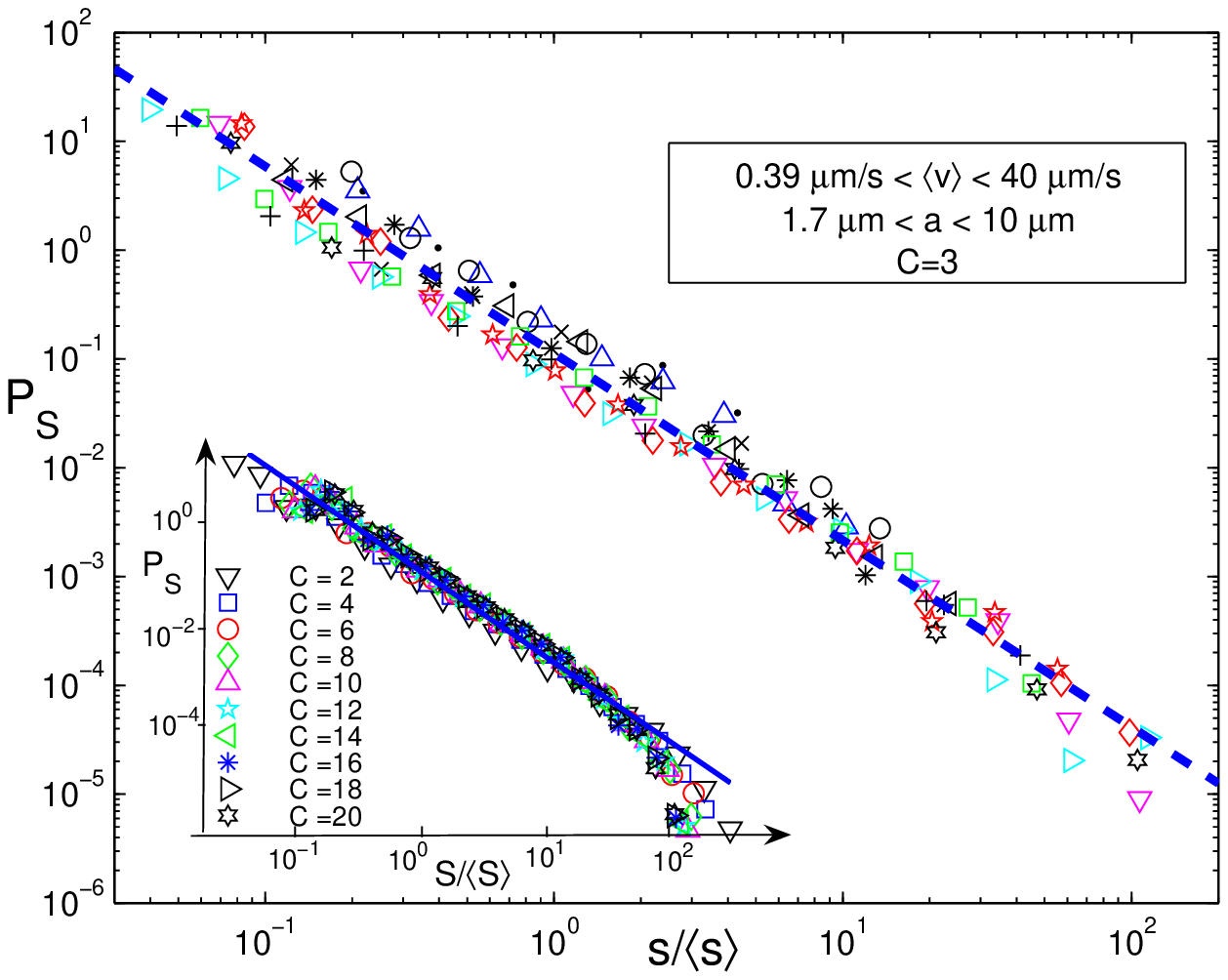}{1.0}{0} \figcap{Fig3}{ Burst size $S$ distribution
 $P(S/\langle S \rangle)$, normalized by the average burst size
 $\langle S \rangle$, for different experimental conditions (the
 various symbols correspond to those on Fig.~\ref{Fig2}). The bursts
 detected for each experiment correspond to clusters of velocities $3$
 times larger than the average crack front speed.  A fit on all the
 data (dashed line) gives a slope equal to $1.71$.  Inset: Normalized
 bursts size distribution $P(S/\langle S \rangle)$ averaged over all
 the different experimental conditions, for a wide range of different
 threshold levels $C$.A fit to all the data,cutting the largest clusters at
 which a cut-off appears due to the lack of statistics (solid line),
 gives a slope equal to $1.67$.}
\end{figure}
Fig.~\ref{Fig1}c shows the spatial distribution of clusters of
different sizes obtained from a thresholded matrix with a  threshold
 level
$C=10$. The white clusters correspond to velocities $10$ times larger
than the average crack front speed $\langle v \rangle$. Then, we can extract from this thresholded  velocity matrix the size 
distribution of the high velocity bursts. The clusters connected to the first and last
front, and thus belonging to the upper and lower white parts are
excluded from the analysis.
  
On Fig.~\ref{Fig3}, the cluster\previous{s} size distribution $P(S/ \langle S
\rangle)$ 
\previous{plot} is shown  for different experiments at a given threshold value $C=3$.  First,
we show a data collapse for all the different experiments performed by
rescaling the clusters size $S$ with the average burst size $\langle S
\rangle $. Moreover, we clearly observe that the burst size
distribution $P(S/ \langle S \rangle)$ follows a power law with an
exponent $\gamma =1.7 $ proving that the burst dynamics occurs on all
length scales.  We have checked that this critical behavio\previous{u}r\rcorrec{,} and 
in particular the exponent $\gamma =1.7 \pm 0.1$\rcorrec{,} is really robust:
normalizing by the average burst size $\langle S \rangle$, we can
rescale all the different distributions corresponding to diverse
experimental conditions and a wide range of threshold level values $2<C<20$
(see inset of Fig.~\ref{Fig3}).

We expect a connection between the spatial scaling of the bursts, 
and the self-affine scaling of the front line itself on large scales.  
To investigate this, we have for each cluster $S$ chosen the smallest rectangular
bounding box enclosing it. The size of the bounding box gives the
length scale $L_y$ of the clusters along the growth direction and the
length scale $L_x$ of the clusters along the average front line
orientation.
\begin{figure}
 \pic{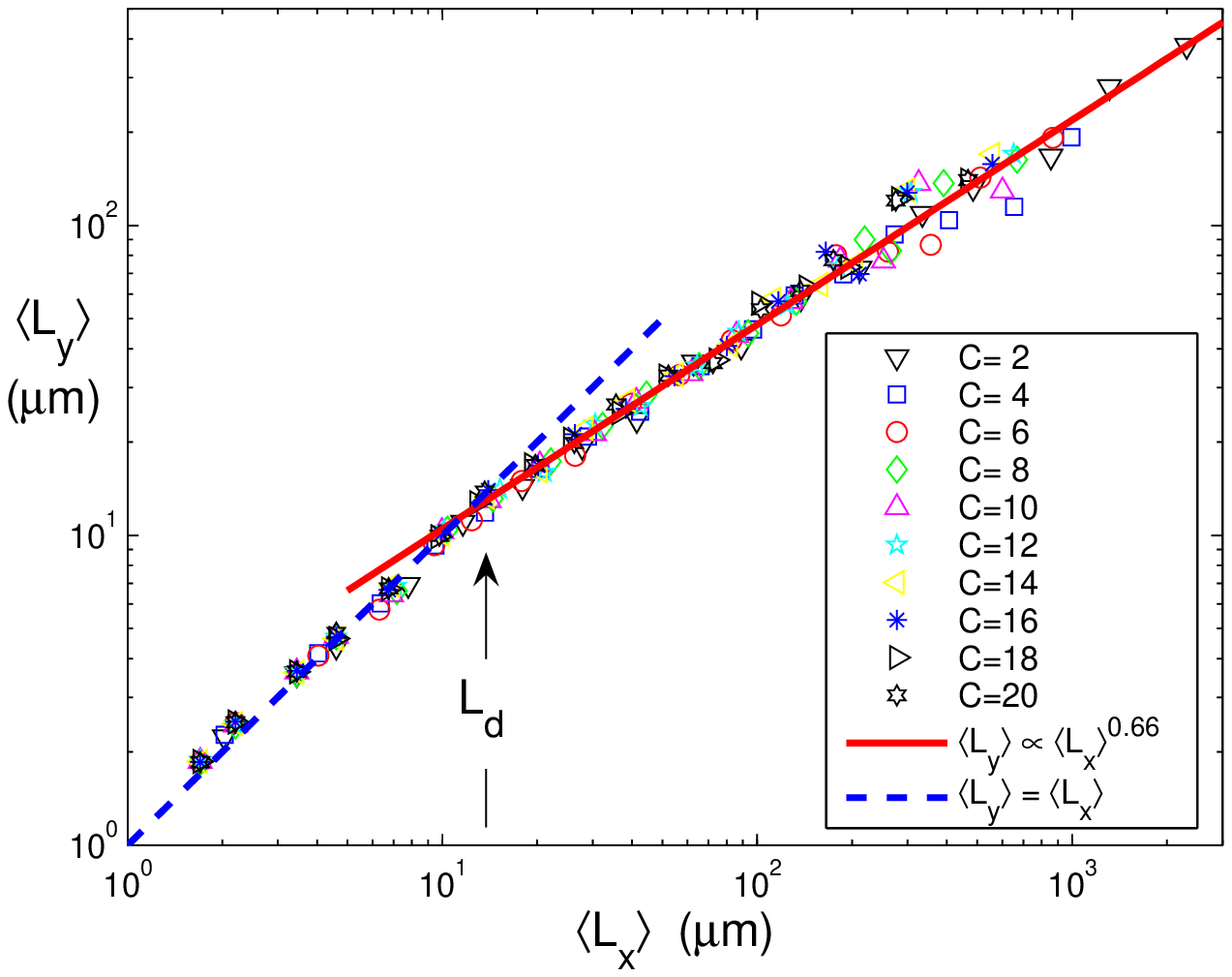}{1.0}{0} \figcap{Fig4}{ Average length scale $\langle
 L_y \rangle$ as function of the \previous{average} length scale $\langle L_x
 \rangle$, for different  threshold levels $C$, averaged over all the
 different experiments performed.  The solid line is a fit to the data
 points for $L_x > 15 \mu m$ and has a slope $H=0.66$, consistent with
 the roughness exponent of the fracture front line. The dotted line
 represents the curve $\langle L_y\rangle = \langle L_x\rangle$ and
 serves as a guide for the eye.}
\end{figure}
Fig.~\ref{Fig4} shows the dependence of the average size $\langle L_y \rangle$
on the  length scale $\langle L_x \rangle$ in a double logarithmic plot, for a wide range
of threshold values $2<C<20$, averaged  over all the different experimental 
conditions (different pixels sizes and average crack front speeds).  
We clearly see that the avalanche
clusters become anisotropic above a characteristic length scale $L_d
\sim 15 \mu m$. This typical size corresponds to the correlation
length for the disorder introduced by the sand-blasting technique
\cite{santucci05b}. Below $L_d$ the local toughness is marked by the
same individual asperity and as a result the thresholded velocity bursts
appear isotropic. A fit to the data points for $L_x > 15 \mu m$ gives
a slope $H=0.66$ consistent with previous independent estimates of the
roughness exponent $\zeta=0.63 \pm 0.03$ for the fracture front
line. This result shows that the system exhibits self-affine scaling
with the same roughness exponent $\zeta$ for the local burst as the
fracture front line and brings a new confirmation of the roughness
exponent found in our experiment, which is different and  higher
than most present theoretical or numerical predictions
\cite{schmittbuhl95b,bouchaud93,roux94,ramanathan98}.

As mentioned in introduction, the scaling behavior of elastic
interfaces in random media is involved in various physical systems.
Actually, the discrepancy between theoretical and measured roughness
exponent has also been reported recently, for contact lines of
helium-4 \cite{Prevost02} and water meniscus
\cite{Moulinet02,Moulinet04} propagating on rough substrates.  It
confirmed that the Joanny-De Gennes model \cite{joanny84} usually
proposed to describe the contact lines dynamics, which leads to the
same kind of equation of motion for crack fronts \cite{Gao89}, is not
sufficient \cite{Moulinet04}. Interestingly the roughness exponent
found respectively equal to $0.56 \pm 0.03$ and $0.52 \pm 0.04$ are close
to the fracture front line roughness. Besides, for the helium-4
meniscus, power law avalanche size distributions have been measured
with exponents $\gamma$ from $0.99$ to $1.3$, depending on the contact
angle, which is different from the exponent $\gamma=1.7$ found in our
experiments. However, avalanches are defined by Prevost et al 
\cite{Prevost02}  based on subtraction between fronts, which is different from 
the present technique, and this $\gamma$ exponent could be sensitive to such definition.

Recent simulations based on a quasi-static model and interpreted as a
stress weighted percolation problem \cite{schmittbuhl03}  give
for the first time consistent results with the experimental roughness
and dynamic exponent \cite{maloy01,santucci05}.  The insensitivity of
the velocity distribution in our experiments on the average velocity
of the front gives support to the quasi-static assumptions used in
these simulations.  However,  the simulated process zone was
not observed in our experiments above the micrometer scale. It should
be mentionned that dynamical effects have also been introduced in a
model \cite{ramanathan98} with a full elastodynamic description where
elastic waves may trigger instabilities and modify the roughness of
the crack front, leading to the value $\zeta=0.5$ \cite{Bouchaud02}.

No theory or simulations so far have investigated the local velocity
or the burst fluctuations.  It will be of great interest to perform
these analysis on the numerical models for a direct comparison with
our experimental work.  This will hopefully clarify the importance of
dynamical effects in modeling the fracture front propagation.

We thank A. Hansen, E.G. Flekk{\o}y, and J.P. Vilotte for fruitful
discussions, and E.L. Hinrichsen at SINTEF-Oslo for his hospitality.
This work was supported by the CNRS/NFR PICS program, the NFR Petromax
and SUP program, and the french programs ACI ``RNCC'' and ACI
``ALEAS''.



\end{document}